\documentclass[prD,showpacs,superscriptaddress,preprintnumbers,twocolumn,amsmath,nofootinbib]{revtex4}
 \usepackage[dvips,final]{graphicx}
  \usepackage{amssymb,amsmath,epsfig,bm,pifont}
   \graphicspath{{./figs/}}
\usepackage{color}
 
  \newcommand{\BluTn}[1]{\textcolor{blue}{#1}}
   \newcommand{\RedTn}[1]{\textcolor{red}{#1}}
\begin{document}
\thispagestyle{empty}
 \date{\today}
  \preprint{\hbox{RUB-TPII-02/2011}}
%\vspace*{-10mm}

\title{Pion-photon transition---the new QCD frontier\\ }
 \author{A.~P.~Bakulev}
  \email{bakulev@theor.jinr.ru}
   \affiliation{Bogoliubov Laboratory of Theoretical Physics, JINR,
                141980 Dubna, Russia\\}

 \author{S.~V.~Mikhailov}
  \email{mikhs@theor.jinr.ru}
   \affiliation{Bogoliubov Laboratory of Theoretical Physics, JINR,
                141980 Dubna, Russia\\}

 \author{A.~V.~Pimikov}
  \email{pimikov@theor.jinr.ru}
   \affiliation{Bogoliubov Laboratory of Theoretical Physics, JINR,
                141980 Dubna, Russia\\}

 \author{N.~G.~Stefanis}
  \email{stefanis@tp2.ruhr-uni-bochum.de}
   \affiliation{Institut f\"{u}r Theoretische Physik II,
                Ruhr-Universit\"{a}t Bochum,
                D-44780 Bochum, Germany}

\begin{abstract}
We perform a detailed analysis of all existing data
(CELLO, CLEO, BaBar) on the pion-photon transition form factor
by means of light-cone sum rules in which we include the
NLO QCD radiative corrections and the twist-four contributions.
The NNLO radiative correction together with the twist-six
contribution are also taken into account in terms of theoretical
uncertainties.
Keeping only the first two Gegenbauer coefficients $a_2$ and $a_4$, we
show that the $1\sigma$ error ellipse of all data up to 9~GeV$^2$
greatly overlaps with the set of pion distribution amplitudes obtained
from nonlocal QCD sum rules---within the range of uncertainties due to
twist-four.
This remains valid also for the projection of the $1\sigma$ error
ellipsoid on the $(a_2,a_4)$ plane when including $a_6$.
We argue that it is not possible to accommodate the high-$Q^2$ tail
of the BaBar data with the same accuracy, despite opposite claims by
other authors, and conclude that the BaBar data still pose a challenge
to QCD.
\end{abstract}
%PACS99 used: Renormalization group evolution of parameters=11.10.Hi
%             Perturbative calculations=12.38.Bx
%             Other nonperturbative calculations in QCD=12.38.Lg
%             Electromagnetic Form Factors=13.40.Gp
%\keywords{Transition form factors,
%          pion distribution amplitude,
%          higher twist,
%          light-cone sum rules,
%          collinear factorization,
%          higher-order radiative corrections,
%          renormalization group evolution}

\maketitle

\section{Introduction}
\label{sec:intro}

The pion-photon transition
$\gamma^{*} + \gamma^{*}\rightarrow\pi^0$
is an example par excellence of an exclusive process that can be
adequately described within QCD by virtue of collinear factorization,
provided both photon virtualities are sufficiently large
(for a review, see \cite{BL89}).
Then, the transition form factor, defined by the correlator of two
electromagnetic currents
\begin{eqnarray}
&&  \int\! d^{4}z\,e^{-iq_{1}\cdot z}
  \langle
         \pi^0 (P)\mid T\{j_\mu(z) j_\nu(0)\}\mid 0
  \rangle
\nonumber \\
&& =
  i\epsilon_{\mu\nu\alpha\beta}
  q_{1}^{\alpha} q_{2}^{\beta}
  F^{\gamma^{*}\gamma^{*}\pi}(Q^2,q^2)
 \label{eq:matrix-element}
\end{eqnarray}
%Eq (1)
with
$Q^2\equiv-q_{1}^2 >0$, $q^2\equiv -q_2^2\geq 0$,
can be recast in leading-twist approximation in the convolution form
\cite{ER80,LB80}
\begin{eqnarray}
  F^{\gamma^{*}\gamma^{*}\pi}(Q^2,q^2)
&\!\!\!\!=\!\!\!\!& \frac{\sqrt{2}}{3}f_\pi\int_{0}^{1}\!dx\,
  T(Q^2,q^2,\mu^2_\text{F},x)\,
\nonumber\\
& &\times \ \varphi^{(2)}_{\pi}(x,\mu^2_\text{F})
  + O\left(\delta^2/Q^{4}\right)\,,
\label{eq:convolution}
\end{eqnarray}
%Eq (2)
where the pion decay constant is $f_\pi=132$~MeV,
$\delta^2$ is the twist-four coupling, and where we assumed
that the photon momenta are sufficiently large
$Q^2, q^2 \gg m_\rho^2$.
This way, the quark-gluon sub-processes, encoded in the hard-scattering
amplitude $T$, can be systematically computed order-by-order within QCD
perturbation theory.
The binding (nonperturbative) effects are separated out and absorbed
into a universal pion distribution amplitude (DA) of twist-two
$\varphi^{(2)}_\pi(x,\mu^2)$,
defined first in \cite{Rad77}.
The variation of $\varphi^{(2)}_\pi(x,\mu^2)$ with the factorization
scale $\mu^2$ is controlled by the
Efremov--Radyushkin--Brodsky--Lepage (ERBL)
renormalization-group evolution equation \cite{ER80,LB80}.
This implies that the pion DA can be expressed in terms of the
Gegenbauer polynomials $C^{3/2}_{n}(2x-1)$ to read
\begin{eqnarray}
  \varphi^{(2)}(x, \mu^2)
= \varphi^\text{as}(x)
  + \sum\nolimits_{n=2,4,\ldots}
  a_{n}\left(\mu^2\right) \psi_{n}(x)\,,
\label{eq:pion-DA.Geg}
\end{eqnarray}
%Eq (3)
where $\psi_{n}(x)\!=\!6x(1-x) C^{3/2}_{n}(2x-1)$
and
$\varphi^\text{as}(x)\!=\!\psi_{0}(x)\!=\!6x(1-x)$ is the asymptotic
pion DA~\cite{ER80,LB80}.

The coefficients $a_{n}$ ($n\geqq 2$) have to be determined within
some nonperturbative model or be extracted from the data,
taking into account evolution effects to account for their
momentum-scale dependence.
In most theoretical analyses they are derived from the moments
$
\langle \xi^{N} \rangle_{\pi}
\equiv
  \int_{0}^{1} dx (2x-1)^{N} \varphi_{\pi}^{(2)}(x,\mu^2)
 $
with the normalization condition
$\int_{0}^{1} dx \varphi_{\pi}^{(2)}(x, \mu^2)=1$.

A process with two photons of high virtuality cannot be easily
measured experimentally.
Experimental information is mostly available for an asymmetric
configuration in which one of the photons is quasi-on-mass-shell
with $q^2\to 0$ \cite{CELLO91,CLEO98,BaBar09}.
In that case, perturbative QCD cannot be directly applied at a
consolidated level because the real photon is emitted at large
distances and has, therefore, a hadronic content which demands
the application of nonperturbative techniques.
Such a method is provided by light-cone sum rules (LCSRs) \cite{Kho99}
(a recent application of this method to the pion-photon transition
form factor is given in \cite{ABOP10}) which augments QCD perturbation
theory in approaching the real photon via a dispersion relation for
$F^{\gamma^*\gamma^*\pi}$ in the variable $q^2$,
keeping the large variable $Q^2$ fixed, viz.,
\begin{equation}
  F^{\gamma^{*}\gamma^{*}\pi}\left(Q^2,q^2\right)
= \int_{0}^{\infty}\!\!ds\,
  \frac{\rho\left(Q^2,s\right)}{s+q^2}\,,
\label{eq:dis-rel}
\end{equation}
%Eq (4)
where the physical spectral density $\rho(Q^2,s)$
approaches at large $s$ the perturbative one:
\begin{equation}
  \rho^\text{PT}(Q^2,s)
=
  \frac{1}{\pi} {\rm Im}F^{\gamma^*\gamma^*\pi}
  \left(Q^2,-s-i\varepsilon\right)\,.
\label{eq:spec-dens-NLO}
\end{equation}
%Eq (5)
The resulting LCSR can be written as follows:
\begin{widetext}
\begin{equation}
  Q^2 F^{\gamma^*\gamma\pi}\left(Q^2\right)
=
  \frac{Q^2}{m_{\rho}^2}
        \int_{x_{0}}^{1}
                        \exp\left(\frac{m_{\rho}^2-Q^2\bar{x}/x}{M^2}
                            \right)
                                   \bar{\rho}(Q^2,x) \frac{dx}{x}
      + \int_{0}^{x_0}
                      \bar{\rho}(Q^2,x) \frac{dx}{\bar{x}}
 \label{eq:LCSR-FF}
\end{equation}
%Eq (6)
\end{widetext}
with the spectral density
$\bar{\rho}(Q^2,x)=(Q^2+s)\rho^\text{PT}(Q^2,s)$
and the abbreviations
$s =\bar{x}Q^2/x$ and
$x_0 = Q^2/(Q^2+s_0)$.
The first term in (\ref{eq:LCSR-FF}) stems from the hadronic content
of a quasi-real photon at low $s\leq s_0$, while the second one
resembles its point-like behavior at higher $s>s_0$.
The hadronic threshold in the vector-meson channel has the value
$s_0=1.5$~GeV$^2$, whereas
$m_{\rho}=0.77$~GeV
\protect\cite{PDG2010}.
The Borel parameter $M^2$ entering the LCSR in (\ref{eq:LCSR-FF})
is not varied aiming to obtain the best stability of the LCSR.
It is actually specified via
$M^2=M_\text{2-pt}^2/\langle{x}\rangle_{Q^2}$
from the two-point QCD sum rule for the $\rho$-meson,
where the corresponding Borel parameter is
$M_\text{2-pt}^2\in[0.5 \div 0.8]$~GeV$^2$.
Here, $\langle{x}\rangle_{Q^2}$ is some average value of $x$
at a fixed scale $Q^2$ in the integration region for
the first integral on the r.h.s. of Eq.\ (\ref{eq:LCSR-FF}), i.e.,
$x_0(Q^2)<\langle{x}\rangle_{Q^2}<1$---see also \cite{Kho99}.

The remainder of the paper is organized as follows.
In the next section, we review the current understanding of
the experimental data on the pion-photon transition form factor,
focusing on the problems related to the unexpected rise of the
BaBar data \cite{BaBar09} beyond 10~GeV$^2$.
In Sec.\ \ref{sec:data-analysis}, we present the salient features of
a calculation of the pion-photon transition form factor using
LCSRs with NLO accuracy and including the twist-four contribution.
The results of our calculation are presented and discussed in
detail in Sec.\ \ref{sec:discussion} in terms of two tables
and several figures, paying particular attention to the intrinsic
theoretical uncertainties stemming from various sources.
Finally, a summary of our main findings is given in Sec.\
\ref{sec:concl}, where we also draw our conclusions.

\section{Status of experimental data}
\label{sec:rev-exp}

Using the method of LCSRs with fixed $M^2=0.7$~GeV$^2$,
Schmedding and Yakovlev (SY) \cite{SY99}
performed a next-to-leading order (NLO) QCD analysis of the data on
$F^{\gamma^*\gamma\pi}$,
obtained by the CLEO Collaboration \cite{CLEO98},
with the inclusion of twist-four contributions, which they modeled
by the asymptotic form, notably, \cite{BF89,Kho99},
\begin{equation}
  \varphi_{\pi}^{(4)}(x,\mu^2)=
  (80/3)\,\delta^2(\mu^2)\,x^2(1-x)^2 \, .
\label{eq:twist-four-DA}
\end{equation}
%Eq (7)
The major outcome of their analysis was that the best agreement with
these high-precision data can be achieved by a pion DA with
the coefficients
$a_{2}^\text{SY}(\mu_\text{SY}^2\equiv5.76~\text{GeV}^2)=0.19$,
$a_{4}^\text{SY}(\mu_\text{SY}^2)=-0.14$,
which after evolution to the reference scale $\mu_1^2\equiv1$~GeV$^2$
\cite{BMS05lat} become
$a_{2}^\text{SY}(\mu_1^2)=0.27$, $a_{4}^\text{SY}(\mu_1^2)=-0.22$.

At the same time, the asymptotic pion DA and the Chernyak--Zhitnitsky
(CZ) model \cite{CZ84} with $a_{2}^\text{CZ}(\mu_1^2)=0.56$ and $a_4=0$
were found to be outside the $2\sigma$ level.\footnote{The value
of $a_{2}^{\rm CZ}$ was originally determined by Chernyak and Zhitnitsky
at the (low) scale $\mu_{\rm CZ}^2=0.25$~GeV$^2$ \cite{CZ84} and reads
$a_{2}^{\rm CZ}(\mu_{\rm CZ}^2)=2/3$.
Its determination at the usual reference scale $\mu_{1}^2=1$~GeV$^2$ is
described in App. B of Ref.\ \cite{BMS02}.}
Later, this type of analysis was refined in \cite{BMS02,BMS03}
by taking into account the correct ERBL evolution of the pion DA to
each measured momentum, including also the variation of the twist-four
coupling
$\delta^2\equiv\delta^2(\mu_1^2)=0.19\pm0.04$~GeV$^2$ in
(\ref{eq:twist-four-DA}).
This way it was shown that the DAs
$\varphi_{\pi}^\text{as}$ and $\varphi_{\pi}^\text{CZ}$
are outside the $3\sigma$ and $4\sigma$ error ellipses, respectively,
while the best fit yields $\chi^2_\text{ndf}=0.47$
(where ndf denotes the number of degrees of freedom),
see for more details in~\cite{BMS03}.

Even more important is the fact that the $1\sigma$ error
ellipse\footnote{%%%
Here and below we denote by $1\sigma$ ellipse (ellipsoid) a confidence
region in the $(a_2, a_4)$ plane that represents a coverage
probability equal to $68.27\%$.}
of the CLEO data strongly overlaps with the admissible values
of the coefficients $a_2$ and $a_4$, determined before in the framework
of QCD sum rules with nonlocal condensates (NLC SRs) \cite{BMS01}.
In particular, we proposed a ``bunch'' of admissible pion DAs
with the central point
$a_{2}^\text{BMS}(\mu_1^2)=0.20$
and
$a_{4}^\text{BMS}(\mu_1^2)=-0.14$
(termed the BMS model) that turns out to be within the $1\sigma$ error
ellipse of the CLEO data \cite{CLEO98}.
These pion DAs
(and similar ones obtained by two of us (BP) in \cite{BP06,AB06parus}
by employing an extended version of the Gaussian model of the nonlocal
QCD vacuum)
have a distinctive feature:
their endpoints at $x=0$ and $x=1$ are strongly suppressed---even
relative to the asymptotic pion DA.
This suppression is due to the finite virtuality
$
 \lambda_{q}^{2}
=
 \langle \bar{q} \left(i g\,\sigma_{\mu \nu}G^{\mu \nu} \right) q
 \rangle / (2\langle \bar{q} q \rangle)=(0.4-0.5)~\text{GeV}^2
$
of the vacuum quarks, which enters the NLC SR for $\varphi_{\pi}^{(2)}$
in terms of the various condensate contributions
(more details can be found in \cite{BMS01,BMS04kg}).

These findings remain valid even if one assumes a much stronger
contribution from the twist-four pion DA, as estimated in
\cite{BMS05lat} using the renormalon model \cite{BGG04}
(see also \cite{Ag05b}).
The calculation of $F^{\gamma^*\gamma\pi}$ beyond the NLO was carried
out in \cite{MS09} by taking into account that part of the
next-to-next-to-leading order (NNLO) correction that is proportional
to the first coefficient $\beta_0$ of the $\beta$ function,
known from the earlier work in \cite{MMP02}.
It turns out that this radiative correction is negative, thus
providing \emph{suppression} of the form factor at the level of
about $8\%$, slowly decreasing with increasing $Q^2$.
To the extent that one's ambition is to comply with the CLEO data on
$\gamma^*\gamma\rightarrow\pi^0$,
it is mandatory to have a pion DA that has its endpoints $x=0,1$
suppressed with a moderate dip at $x=1/2$, just as the BMS model
extracted from NLC SRs.
It is worth mentioning that this model gives rise to a scaled form
factor
$Q^2 F^{\gamma^*\gamma\pi}(Q^2)$
that approaches the asymptotic limit $\sqrt{2}f_\pi$,
predicted by perturbative QCD, from \emph{below},
while at low momenta it also yields reasonable agreement
with the CELLO data \cite{CELLO91}.\footnote{%%%
Note that at low momenta close to $Q^2\approx 1$~GeV$^2$,
the twist-two contribution becomes comparable in size
with the twist-four term.}
Therefore, it came as a surprise that the data on
$\gamma\gamma^\ast\to \pi^0$, presented by the BaBar Collaboration
in 2009 \cite{BaBar09}, indicate above approximately $10$~GeV$^2$
a marked growth of the form factor with $Q^2$
(see upper points (labeled by open crosses)
in Fig.\ \ref{fig:BaBar-eta-fig18}
taken from \cite{BaBar11-BMS}).
In fact, the BaBar Collaboration fitted their data with the expression
\begin{equation}
  Q^2 F^{\gamma^{*}\gamma\pi}
\sim
  \left(
        \frac{Q^2}{10~{\rm GeV}^2}
  \right)^\beta\,,
\label{eq:power}
\end{equation}
%Eq (8)
where $\beta=0.25 \pm 0.02$, i.e.,
$F^{\gamma^{*}\gamma\pi} \sim Q^{-3/2}$.
This behavior of the form factor is incompatible with the collinear
factorization and exceeds the asymptotic prediction considerably
up to the highest measured momentum value just below 40~GeV$^2$.
Moreover, the high-$Q^2$ BaBar tail contradicts the CLEO data
which are well-described by the dipole fit \cite{CLEO98}
\begin{equation}
  Q^{2}F^{\gamma^{*}\gamma\pi}
\sim
  \frac{Q^2}{Q^2+\Lambda^2}
\label{eq:dipole}
\end{equation}
%Eq (9)
with
$\Lambda\approx 776$~MeV.
This scale behavior becomes more puzzling by comparing the
$\pi^0$ BaBar data with the predictions extracted by the BaBar
Collaboration from their analysis of the two-photon $\eta$
and $\eta^\prime$ decays \cite{BaBar11-BMS}---Fig.\
\ref{fig:BaBar-eta-fig18}.
While the direct $\pi^0$ data grow with $Q^2$ and necessitate a wide DA
with strong contributions from the endpoint regions $x=0, 1$,
the converted $\eta-\eta^\prime$ data are in perfect agreement
with the prediction obtained with the endpoint-suppressed BMS pion DA
(solid line in Fig.\ \ref{fig:BaBar-eta-fig18}).
To describe both sets of the BaBar data with the \emph{same} accuracy
looks like a catch-22-situation because it demands an impossible
compromise for the structure of the pion DA
(more on this and its implications later).

Many theorists have attempted during the past two years to explain
the rise of the BaBar data using various schemes and drawing
incongruent conclusions.
We may group these theoretical proposals into four categories:

%%%%%%%%%%%%%%%%%%%%%%%%%%%%%%%%%%%%%%%%%%%%%%%%%%%%%%%%%%%%%%%%%%%%%%% Figure 1
\begin{figure}[t!]
\includegraphics[width=0.45\textwidth]{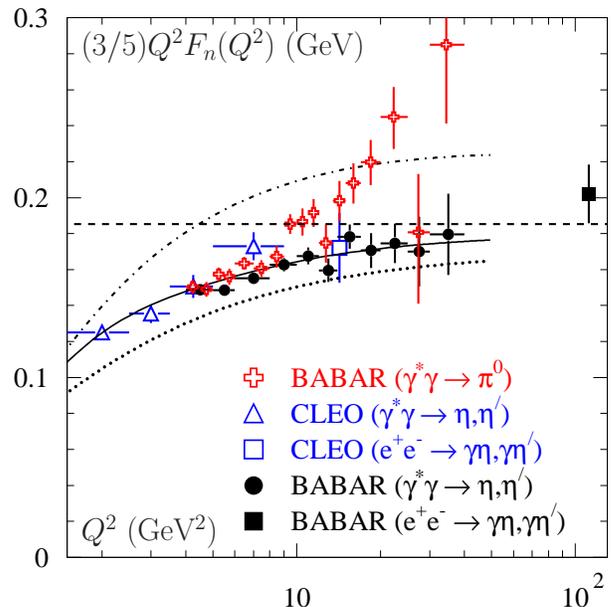}
\vspace{-5mm}
\caption{The $\gamma\gamma^\ast\to |n\rangle$ transition form factor
multiplied by $3/5$ in comparison with
$Q^2 F^{\gamma\gamma^\ast\pi}(Q^2)$ \cite{BaBar09}.
The dashed line indicates the asymptotic limit for
$Q^2 F^{\gamma\gamma^\ast\pi}(Q^2)$.
The dotted, dashed-dotted, and solid curves show
the predictions from \cite{BMS02,BMS03}
for the asymptotic,
the CZ,
and the BMS DA, respectively
(figure taken from ~\cite{BaBar11-BMS}).
\label{fig:BaBar-eta-fig18}
\vspace{-7mm}}
\end{figure}
%%%%%%%%%%%%%%%%%%%%%%%%%%%%%%%%%%%%%%%%%%%%%%%%%%%%%%%%%%%%%%%%%%%%%%%

(i) \emph{Abandon collinear factorization}
\\
An increase of the form factor with $Q^2$ can be achieved by employing
a flat-top pion DA, as proposed by Radyushkin \cite{Rad09} and by
Polyakov \cite{Pol09}---which used, however, different schemes.
The key element of these approaches is a nonvanishing pion DA
at the endpoints $x=0,1$.
Such a pion DA ascribes the rise of the large-$Q^2$ tail of the BaBar
data to a logarithmic increase of the form factor induced by the
flat-top DA.
Note, incidentally, that as shown in \cite{MPS10}, a $\ln Q^2$ behavior
can actually be imitated by including just the first three Gegenbauer
polynomials without demanding the flatness of the pion DA.
However, one has to pay a high price for this agreement:
already the NLO contribution at the default normalization scale
$\mu^2=Q^2$ turns out to be huge for the flat-top DA, forcing a much
lower normalization scale below $\Lambda_\text{QCD}$ and making it
virtually impossible to include evolution effects which are at the
heart of QCD.

(ii) \emph{Deny rise of the form factor}
\\
Very recently, Agaev et al.\ \cite{ABOP10} argued that the trend of
the high-$Q^2$ tail of the $\pi^0$ BaBar data can be reproduced
within QCD using LCSRs.
Their analysis of the form factor includes radiative corrections at
the NLO level, twist-four contributions, and, for the first time,
also a twist-six term which, though it has a positive sign,
turns out to be small for their choice of the Borel parameter
$M^2=1.5$~GeV$^2$ to provide any serious enhancement.\footnote{%%%%
This fact is emphasized by these authors as being rather crucial
in selecting that large value of the Borel parameter.}
The distinctive feature of their models for the pion DA is a large
and positive coefficient $a_4$ to which they attribute the rise of
the form factor at high $Q^2$, while still higher coefficients $a_{n}$
up to $n=12$ have smaller values.
The best agreement with the $\pi^0$ BaBar data \cite{BaBar09}
is provided by a modified flat-like DA with $a_2=0.13$, which has a
large tail of Gegenbauer coefficients, while the other two considered
models are given by truncated expansions with $a_{n}$ up to $n=8$
and provide form-factor predictions that almost scale with $Q^2$,
but having a normalization exceeding considerably the asymptotic limit
of $\sqrt{2}f_\pi$.
Even the model with $a_{n}$ coefficients up to $n=12$, which become
successively operative with increasing $Q^2$, cannot simulate the
increase of the form factor sufficiently well in comparison
with the BaBar fit (\ref{eq:power}) used in \cite{BaBar09}---provided
one takes this fit seriously.
At the same time, as one may appreciate from
Fig.\ \ref{fig:BaBar-eta-fig18},
all these models overestimate the BaBar data \cite{BaBar11-BMS}
on the $\gamma\gamma^*\to\eta(\eta^\prime)$
transition form factor significantly.
Taking into account the NNLO radiative corrections, would somewhat
reduce the magnitude of the form factor by a few \% \cite{MS09},
but incorporating this effect would also shift the prediction further
away from the $\gamma^*\gamma\to\pi^0$ BaBar data---the main goal.
Thus, the conclusion drawn in \cite{ABOP10} that one may achieve
simultaneous agreement with all data on
$F^{\gamma^*\gamma\pi}$ using LCSRs within QCD seems rather
perfunctory.
At the same time, the fact that the twist-six term turns out to have
a positive sign renders Chernyak's recent claim in \cite{Che09} that
one can describe all data using the CZ DA supplemented by a power
correction $1/Q^6$ with a huge negative coefficient $-$(1.2~GeV$^2$)$^2$
rather speculative.\footnote{%%
However, other power corrections of different origin with negative sign
cannot be excluded, but can hardly be accommodated within QCD.}

(iii) \emph{Contextual explanations}
\\
Various authors (e.g., \cite{Dor09}) have presented calculations of
$Q^2 F^{\gamma^*\gamma\pi}$
within the context of particular models---appealing, for instance, to
the triangle anomaly---and have indeed obtained a logarithmic increase,
emulating this way the large-$Q^2$ tail of the BaBar data.
The problem with such approaches is that they involve (nonperturbative)
context-dependent mass scales that cannot be derived from QCD and are
specific for this particular process.

(iv) \emph{$\bm{k}_\perp$ factorization}
\\
Some authors try to explain the increase of the form factor by
retaining the partonic $\bm{k}_\perp$ degrees of freedom unintegrated
and employing a modified $\bm{k}_\perp$ factorization approach, see for
instance \cite{LiMi09,WH10,Kro10sud,BCT11}.
However, it is well-known from previous research \cite{KR96,SSK99}
that Sudakov effects and $\bm{k}_\perp$-generated power corrections
give rather suppression than enhancement, albeit self-reinforcement
is also possible \cite{KS01}.
This subject is, however, outside the scope of the present
investigation; it will be discussed in detail elsewhere.

\section{Data analysis}
\label{sec:data-analysis}

The strategy of our investigation is to use LCSRs in order to
process all available experimental data on the pion-photon transition
form factor from different experiments \cite{CELLO91,CLEO98,BaBar09},
making some technical improvements relative to our previous works
in \cite{BMS02,BMS03,BMS05lat,MS09}, and extending the analysis
to the level of three Gegenbauer coefficients $a_2, a_4, a_6$.
The improvements and the applied algorithms are explained in detail
below.

To compute the form factor, we employ the LCSR (\ref{eq:LCSR-FF}).
In order to resolve the $\rho$ and $\omega$ meson resonances,
we employ not a simple $\delta$-function ansatz,
but a finite-width Breit--Wigner form~\cite{Kho99,MS09}.
Adopting Khodjamirian's general arguments \cite{Kho99}, we relate
the Borel parameter $M^2$ in the LCSR (\ref{eq:LCSR-FF}) to the
Borel parameter $M_\text{2-pt}$ entering the two-point QCD sum rule
for the $\rho$ meson~\cite{SVZ}.
Hence, instead of using the fixed value $M^2=M_0^2=0.7$~GeV$^2$,
as in our previous works \cite{BMS02,BMS03,BP06,AB06parus,MS09},
we write
$M^2(Q^2)=M_0^2/\langle x\rangle_{Q^2}$,
where the average $\langle x\rangle_{Q^2}$
is calculated by means of the Borel exponential
$\exp\left[Q^2\bar{x}/(M_0^2\,x)\right]$
in (\ref{eq:LCSR-FF}).
This prescription amounts to
$M^2(Q^2)\in[0.7 \div 0.9]$~GeV$^2$.

%%%%%%%%%%%%%%%%%%%%%%%%%%%%%%%%%%%%%%%%%%%%%%%%%%%%%%%%%%%%%%%%%%%%%%% FIGURE 2
%\begin{widetext}
\begin{figure*}[t!]
\centerline{\includegraphics[width=0.4\textwidth]{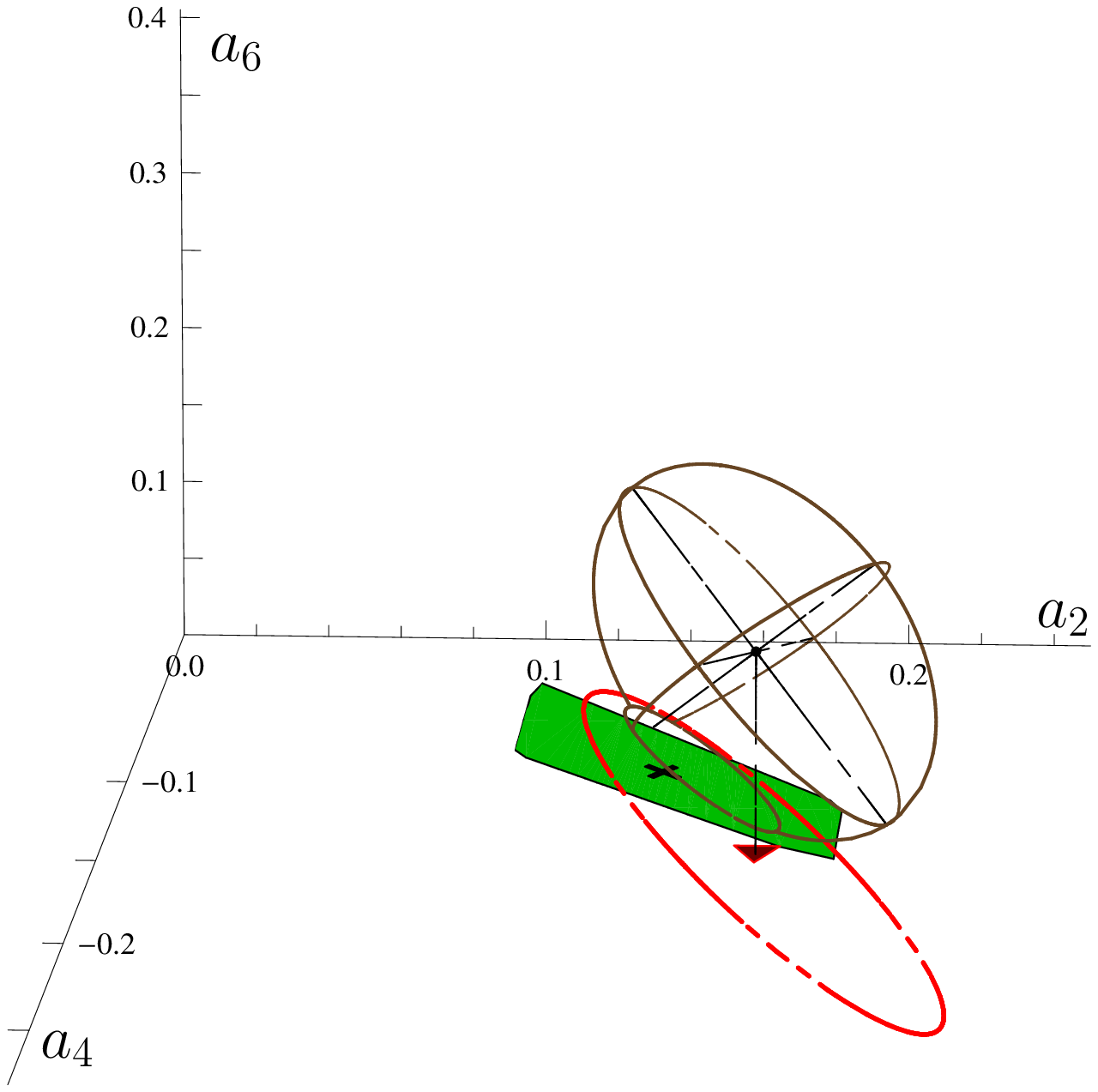}~~%
~~~~~~~~~~~~\includegraphics[width=0.4\textwidth]{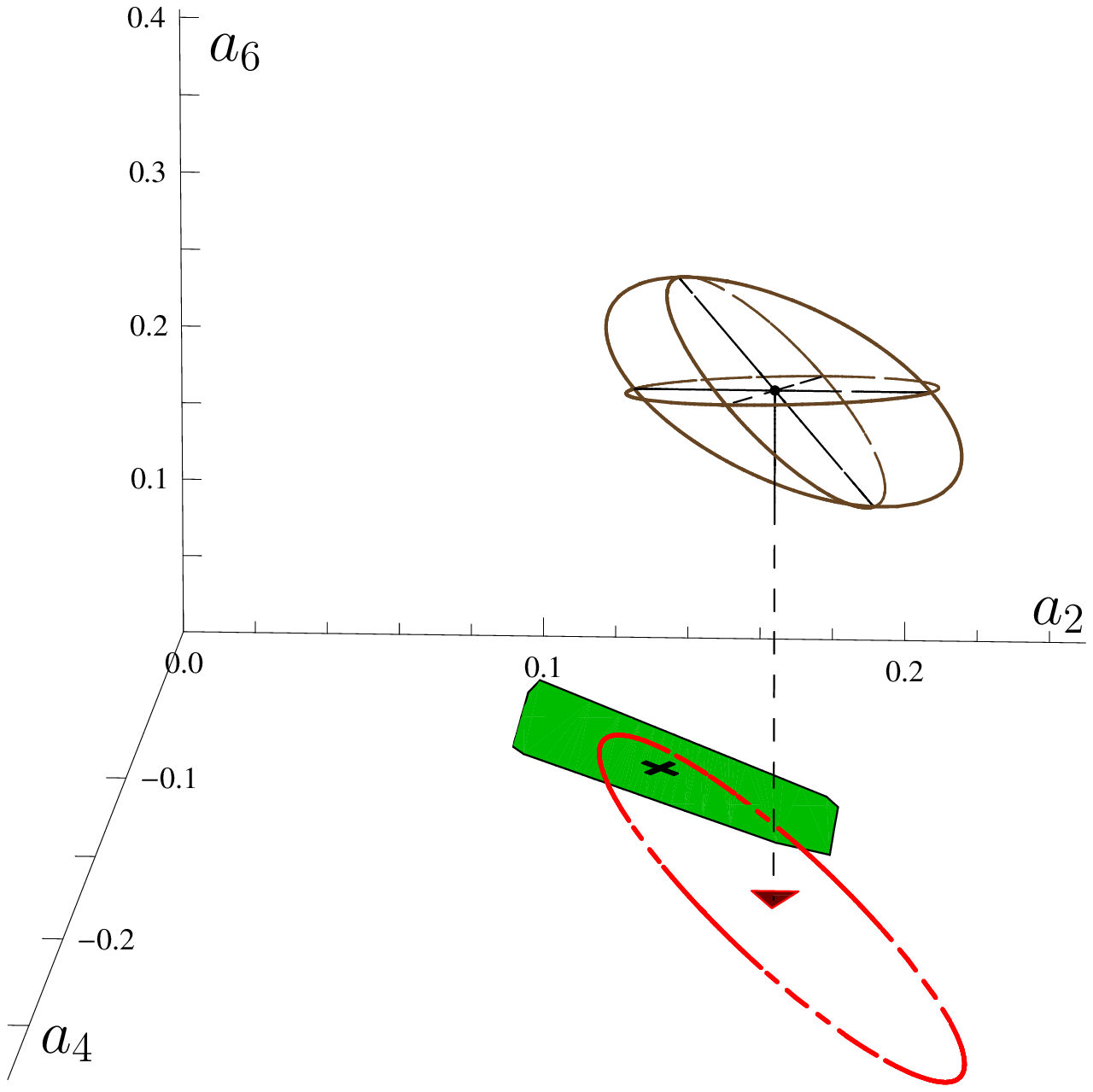}}
\caption{(color online). 3D graphics of $1\sigma$ error ellipsoids in the
space spanned by the Gegenbauer coefficients $a_n$ $(n=2,4,6)$ of all
data \protect\cite{CELLO91,CLEO98,BaBar09} on the pion-photon
transition form factor in the range $[1 \div 9]$~GeV$^2$ (left panel)
and $[1 \div 40]$~GeV$^2$ (right panel) processed with the help of
LCSRs (details in the text).
The projection of the $1\sigma$ ellipsoid on the plane $(a_2,a_4)$
is represented, in both panels, by the larger ellipse in red color.
The smaller ellipse denotes the cross section of the ellipsoid with
the $(a_2,a_4)$ plane.
The shaded (green) rectangle encloses the region of those $a_2,a_4$
pairs that are allowed by NLC SRs
\protect\cite{BMS01}, whereas its middle point ({\footnotesize\ding{54}})
marks the BMS pion DA.
All results are shown at the scale $\mu_\text{SY}^2$ after NLO
evolution.
\protect\label{fig:3D-graphs}}
\end{figure*}
%\end{widetext}
%%%%%%%%%%%%%%%%%%%%%%%%%%%%%%%%%%%%%%%%%%%%%%%%%%%%%%%%%%%%%%%%%%%%%%%

The key ingredients of our data-analysis procedure are the following:
\\
(i) The NLO gluon radiative corrections in the spectral density are
taken into account using a corrected expression with respect to
Eq.\ (3.12) in \cite{MS09} which does not contain the error pointed
out in \cite{ABOP10}.
Note that this error does not affect the results of our previous
calculations with LCSRs in \cite{BMS02,BMS03,BMS05lat,MS09}.
We adopt the so-called default renormalization-scale setting and
identify the factorization and the renormalization scale with the
large photon virtuality $Q^2$.
This avoids the appearance of (large) logarithms of these scales.
\\
(ii) We take into account the twist-four contribution,
allowing for a significant variation of the parameter
$\delta^2=0.19$~GeV$^2$
in the range 0.15~GeV$^2$ to $0.23$~GeV$^2$,
referring the reader for a detailed discussion to our previous analysis
in \cite{BMS03}.
As already mentioned, the use of a nonasymptotic form for
$\varphi_{\pi}^{(4)}$
would not affect the results significantly \cite{BMS05lat,Ag05b}.
The parameter $\delta^2(\mu^2)$ is evolved with $\mu^2$ according
to the one-loop renormalization-group equation, whereas other
perturbative corrections to the twist-four contribution
are ignored---expected to be small.
\\
(iii) The evolution of the coefficients $a_n$ is taken into account at
the NLO level of accuracy using the QCD scale parameters
$\Lambda_\text{QCD}^{(3)}=370$~MeV and
$\Lambda_\text{QCD}^{(4)}=304$~MeV,
consistent with the NLO estimate
$\alpha_s(M_Z^2)=0.118$ \cite{PDG2010}.
\\
(iv) \label{page4}
The inclusion of the NNLO$_{\beta_0}$ radiative correction to the
form factor, calculated in \cite{MMP02} and used in the spectral
density in \cite{MS09}, as well as the twist-six term, computed for
the first time in \cite{ABOP10}, are taken into account implicitly
within the theoretical uncertainties.
Their explicit evaluation is relegated to a future dedicated
investigation.
This treatment of the two contributions in terms of uncertainties
is justified by the fact that for the average value of
$M^2(Q^2)\sim 0.75$~GeV$^2$
the net result appears to be small, decreasing with $Q^2$ from
$+0.005$ at $Q^2=1$~GeV$^2$---where the twist-six term prevails---down
to $-0.003$ at $Q^2=40$~GeV$^2$---where the NNLO correction
becomes stronger.
Note that in the calculation of the NNLO term only the convolution of
the hard-scattering amplitude $T_\beta$ with the asymptotic DA
$\varphi^\text{as}$ is taken into account
(see for more details in \cite{MS09}).
It is worth mentioning that this behavior is sensitive to the
choice of the Borel scale adopted in our work.
Would we use instead the value $M^2=1.5\pm0.5$~GeV$^2$,
advocated for in \cite{ABOP10}, then the twist-six term would be much
smaller and the net result would be everywhere negative and almost
constant:
$\approx-0.004$.

%%%%%%%%%%%%%%%%%%%%%%%%%%%%%%%%%%%%%%%%%%%%%%%%%%%%%%%%%%%%%%%%%%%%%%% FIGURE 3
%\begin{widetext}
\begin{figure*}[t]
 \centerline{\includegraphics[width=0.45\textwidth]{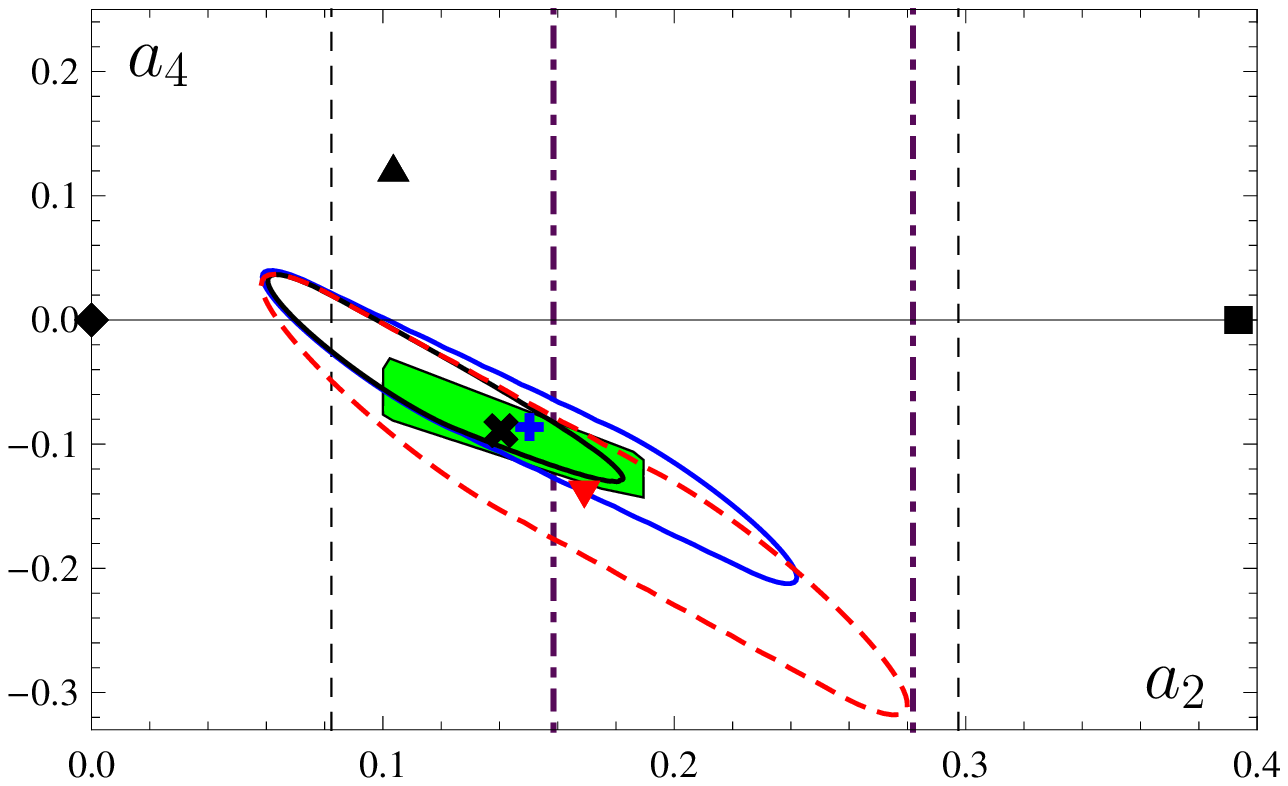}%
~~~~~~~~~~~~~\includegraphics[width=0.45\textwidth]{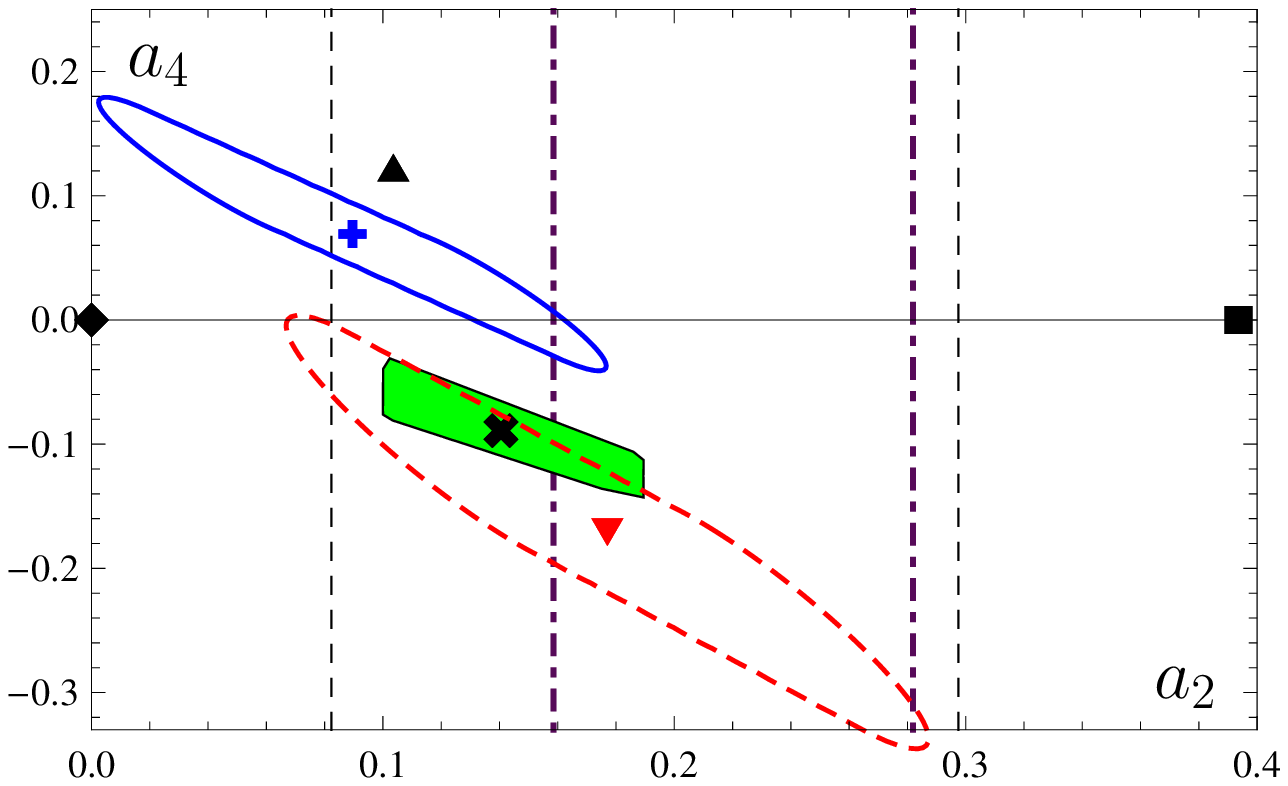}}
\caption{(color online).  Distorted $1\sigma$ error ellipses
of the $F^{\gamma^*\gamma\pi}(Q^2)$ data in the range $[1 \div 9]$~GeV$^2$
(left panel) and $[1 \div 40]$~GeV$^2$ (right  panel) from various experiments
\protect\cite{CELLO91,CLEO98,BaBar09,BaBar11-BMS}
using different data-analysis procedures.
These ellipses (see for explanations in the text) are the
result of unifying different ellipses which correspond to the
twist-four parameter varying in the range
$\delta^2=0.15 \div 0.23$~GeV$^2$.
The slanted rectangle in green color encloses the area of $a_2$ and
$a_4$ values determined by NLC SRs \protect\cite{BMS01}, with the
BMS pion DA being marked by {\footnotesize\ding{54}}.
The middle points of the ellipses are also labeled
(\BluTn{\footnotesize\ding{58}} and \RedTn{\footnotesize\ding{116}})
together with the asymptotic DA ({\footnotesize\ding{117}}),
the CZ DA ({\footnotesize\ding{110}}), and Model III from
\protect\cite{ABOP10} ({\footnotesize\ding{115}}).
The vertical lines indicate the range of $a_2$ values determined by
two lattice simulations: \cite{Lat06}---dashed lines;
\cite{Lat10}---dashed-dotted (blue) lines.
The Borel parameter $M^2(Q^2)$ is varied with $Q^2$ in both panels
as explained in the text.
All results are shown at the scale $\mu_\text{SY}^2$.
\protect\label{fig:data-ellipses}}
\end{figure*}
%\end{widetext}
%%%%%%%%%%%%%%%%%%%%%%%%%%%%%%%%%%%%%%%%%%%%%%%%%%%%%%%%%%%%%%%%%%%%%%%

\section{Discussion of results}
\label{sec:discussion}

To be able to make precise estimates of the influence of the high-$Q^2$
tail of the BaBar data on the form factors, we use two different data
sets: one which collects all available data from three experiments
\cite{CELLO91,CLEO98,BaBar09} for the $Q^2$ region between 1 and 9
GeV$^2$ (termed for convenience ``CLEO regime'') and a second set
which includes as well the high-$Q^2$ BaBar data up to about
$40$~GeV$^2$.
These results are displayed in Fig.\ \ref{fig:3D-graphs},
left and right panel, respectively.
In practice, this analysis procedure allows us to understand the role
and relative strength of the coefficient $a_6$ in determining the
characteristics of the $1\sigma$ error ellipsoid of each data set
selection, shown in this figure by its principal axes and
corresponding ellipses.

The middle point $(0.17,-0.14,0.12)$ of the $1\sigma$ error ellipsoid
in the left panel of Fig.\ \ref{fig:3D-graphs}, resulting from
analyzing the data in the CLEO regime, corresponds to a rather good
$\chi^2_\text{ndf}\approx0.4$.
This ellipsoid is stretched along the axis $a_6$ and has a sizable
intersection with the $(a_2,a_4)$ plane, shown in the figure as a
smaller ellipse inside the green rectangle.
In contrast, the inclusion of the high-$Q^2$ BaBar data (right panel)
forces the ellipsoid away from that plane to a new middle point
$(0.18,-0.17,0.31)$
with a worse value $\chi^2_\text{ndf}\approx 1$.
Observe that the projection of the ellipsoid on the $(a_2,a_4)$
plane remains almost unchanged, while its middle point
\RedTn{\footnotesize\ding{116}} moves slightly away from its previous
location, albeit it still resides inside the region of the negative
values of $a_4$.
The fact of the matter is that both error ellipses mentioned above
cover a quite large region of those values of the coefficients
$a_2$ and $a_4$
which have been determined in \cite{BMS01}
with the help of NLC SRs---slanted (green) rectangle
in Fig.\ \ref{fig:3D-graphs}.
In particular, the position of the BMS pion DA in the $(a_2,a_4)$ plane
is just inside both ellipses for the CLEO regime with respect to all
data (left panel) and still inside the 2D projection when one takes
into account also the BaBar data up to $40$~GeV$^2$ (right panel).

To further quantify these considerations, we perform a more detailed
data analysis with the focus on the $(a_2,a_4)$ plane and show
the results in Fig.\ \ref{fig:data-ellipses}.
The left panel refers to the data selection for the $Q^2$ values
in the CLEO regime, while the right panel shows the analogous findings
for the second case when we implement the analysis by including all
BaBar data up to the highest measured value of $Q^2$.
To this end, we calculate the $1\sigma$ error ellipses by allowing the
parameter $\delta^2$ to vary by $20\%$ around the value 0.19~GeV$^2$.
All ellipses obtained this way are then merged together into a single
distorted $1\sigma$ error ellipse shown in Fig.\
\ref{fig:data-ellipses} for different situations.
These are the following:
The largest ellipse---dashed line in red color with the middle point
\RedTn{\footnotesize\ding{116}}---represents the result of combining
the projections on the plane $(a_2,a_4)$ of the 3D data analysis.
The smaller ellipse (solid blue line) shows the analogous result
of a 2D analysis by means of $a_2$ and $a_4$.
Its middle point \BluTn{\footnotesize\ding{58}}
with the coordinates $(0.15,-0.09)$
and $\chi^2_\text{ndf} \approx 0.5$,
almost coincides with the middle point {\footnotesize\ding{54}} of the
area of values determined by NLC SRs.
Finally, the smallest ellipse (thick line), entirely enclosed by the
previous one, is obtained by combining the intersections with the
$(a_2,a_4)$ plane of all ellipsoids generated by the variation around
the central value of $\delta^2$.
The ellipsoid which corresponds to the central value
$\delta^2=0.19$~GeV$^2$
is shown in the left panel of Fig.\ \ref{fig:3D-graphs}.

The locations in the $(a_2,a_4)$ plane of some characteristic
pion DAs are also indicated in Fig.\ \ref{fig:data-ellipses};
notably,
the asymptotic DA ({\footnotesize\ding{117}}),
the CZ model ({\footnotesize\ding{110}}),
and also the projection of Model III from \cite{ABOP10}
({\footnotesize\ding{115}}).
It is worth emphasizing that the slanted (green) rectangle
that contains those values of $a_2$ and $a_4$ derived from
NLC SRs \cite{BMS01} lies almost completely within both larger error
ellipses---greatly overlapping with the smallest one as well.
In particular, the BMS model DA {\footnotesize\ding{54}}
turns out to be inside of all $1\sigma$ error ellipses.
Hence, the results of the 2D and 3D data analyses are in good mutual
agreement, while we observe no tension between them and the theoretical
predictions obtained from NLC SRs \cite{BMS01}---at least at the level
of $\chi^2_\text{ndf}\leq0.5$.
What is more, all calculated $1\sigma$ error ellipses are complying
fairly well with the boundaries for $a_2$ provided by two independent
lattice simulations.
The older estimate from \cite{Lat06} is indicated in both panels of
this figure by the vertical dashed lines, whereas the very recent
results given in Ref.\ \cite{Lat10} (Table XII there) are denoted by
the dashed-dotted vertical lines in blue color.

%%%%%%%%%%%%%%%%%%%%%%%%%%%%%%%%%%%%%%%%%%%%%%%%%%%%%%%%%%%%%%%%%%%%%%%
%%%%%%%%%%%%%%%%%%%%%%%%%%%%%%%%%%%%%%%%%%%%%%%%%%%%%%%%%%%%%% TABLE I
%%%%%%%%%%%%%%%%%%%%%%%%%%%%%%%%%%%%%%%%%%%%%%%%%%%%%%%%%%%%%%%%%%%%%%%
\begin{table*}[t!]
\begin{center}
\begin{ruledtabular}
\caption{Results for the scaled pion-photon transition form factor
calculated in this work (last column)
in comparison with the values measured at the same momentum scales
by the CLEO \protect\cite{CLEO98} and
the BaBar \protect\cite{BaBar09,BaBar11-BMS} Collaborations,
both for the $\gamma^*\gamma\to\pi^0$
and the $\gamma^*\gamma\to |n\rangle$ decays.
The last entry at the timelike momentum
$\tilde{Q}^2=112$~GeV$^2$ is taken from \protect\cite{Aub2006}.
The origin of the theoretical uncertainties in the last column
is explained in the text. \label{tab:ff-values-table-1}}
\smallskip
\smallskip
\begin{tabular}{cccccc}
CLEO/BaBar      & CLEO/BaBar
                        & CLEO                     & BaBar             & BaBar ($\eta,\eta'$) & This work
\\
 $Q^2$ {\small interval}
            & $\tilde{Q}^2$
                    & $\tilde{Q}^2 |F_\text{CLEO}^{\gamma^*\gamma\pi^0}(\tilde{Q}^2)|$
                                                 & $\tilde{Q}^2 |F_\text{BaBar}^{\gamma^*\gamma\pi^0}(\tilde{Q}^2)|$
                                                                       & $(3/5)\tilde{Q}^2 |F_\text{BaBar}^{\gamma^*\gamma n}(\tilde{Q}^2)|$
                                                                                            & $\tilde{Q}^2 |F^{\gamma^*\gamma\pi^0}(\tilde{Q}^2)|$
\\ \phantom{.}
 [GeV$^2$]  & [GeV$^2$]
                    & [0.01 $\times$ GeV]        & [0.01 $\times$ GeV] &[0.01 $\times$ GeV] & [0.01 $\times$ GeV]
\\
\hline
 1.5 -- 1.8 &  1.64 & 12.1$\,\pm\,0.8\,\pm\,$0.3 &  --                 &   --               & 11.14$_{-1.02}^{+1.06}$$_{-0.3}^{+0.5}$  \\
 1.8 -- 2.0 &  1.90 & 11.7$\,\pm\,0.7\,\pm\,$0.3 &  --                 &   --               & 12.04$_{-1.03}^{+1.09}$$_{-0.3}^{+0.5}$  \\
 2.0 -- 2.2 &  2.10 & 13.8$\,\pm\,0.8\,\pm\,$0.3 &  --                 &   --               & 12.58$_{-1.01}^{+1.09}$$_{-0.3}^{+0.5}$  \\
 2.2 -- 2.4 &  2.30 & 12.7$\,\pm\,0.9\,\pm\,$0.3 &  --                 &   --               & 13.03$_{-0.99}^{+1.07}$$_{-0.3}^{+0.5}$  \\
 2.4 -- 2.6 &  2.50 & 13.5$\,\pm\,1.0\,\pm\,$0.3 &  --                 &   --               & 13.4 $_{-0.95}^{+1.04}$$_{-0.3}^{+0.5}$  \\
 2.6 -- 2.8 &  2.70 & 15.1$\,\pm\,1.1\,\pm\,$0.4 &  --                 &   --               & 13.71$_{-0.92}^{+1.01}$$_{-0.3}^{+0.5}$  \\\hline
 2.8 -- 3.1 &  2.94 & 13.7$\,\pm\,1.2\,\pm\,$0.3 &  --                 &   --               & 14.02$_{-0.90}^{+0.99}$$_{-0.3}^{+0.5}$  \\
 3.1 -- 3.5 &  3.29 & 14.5$\,\pm\,1.2\,\pm\,$0.4 &  --                 &   --               & 14.38$_{-0.88}^{+0.96}$$_{-0.3}^{+0.5}$  \\
 3.5 -- 4.0 &  3.74 & 13.2$\,\pm\,1.4\,\pm\,$0.3 &  --                 &   --               & 14.73$_{-0.85}^{+0.92}$$_{-0.3}^{+0.5}$  \\
 4.0 -- 4.5 &  4.24 & 13.4$\,\pm\,1.5\,\pm\,$0.3 & 15.04 $\pm$ 0.39    &   --               & 15.02$_{-0.82}^{+0.87}$$_{-0.3}^{+0.5}$  \\
 4.0 -- 5.0 &  4.44 & --                         &  --                 & 14.89$\,\pm\,$0.26 & 15.12$_{-0.80}^{+0.86}$$_{-0.3}^{+0.5}$  \\
 4.5 -- 5.0 &  4.74 & 15.4$\,\pm\,1.7\,\pm\,$0.4 & 14.91$\,\pm\,$0.41  &   --               & 15.25$_{-0.79}^{+0.83}$$_{-0.3}^{+0.5}$  \\\hline
 5.0 -- 5.5 &  5.24 & 14.5$\,\pm\,1.8\,\pm\,$0.4 & 15.74$\,\pm\,$0.39  &   --               & 15.43$_{-0.76}^{+0.80}$$_{-0.3}^{+0.5}$  \\
 5.0 -- 6.0 &  5.45 & --                         &  --                 & 14.86$\,\pm\,$0.27 & 15.49$_{-0.75}^{+0.79}$$_{-0.3}^{+0.5}$  \\
 5.5 -- 6.0 &  5.74 & 15.5$\,\pm\,2.2\,\pm\,$0.4 & 15.60$\,\pm\,$0.45  &   --               & 15.58$_{-0.74}^{+0.77}$$_{-0.3}^{+0.5}$  \\
 6.0 -- 7.0 &  6.47 & 14.8$\,\pm\,2.0\,\pm\,$0.4 & 16.35$\,\pm\,$0.36  &   --               & 15.76$_{-0.71}^{+0.74}$$_{-0.3}^{+0.5}$  \\
 6.0 -- 8.0 &  6.85 & --                         &  --                 & 15.52$\,\pm\,$0.28 & 15.84$_{-0.70}^{+0.73}$$_{-0.3}^{+0.5}$  \\
 7.0 -- 8.0 &  7.47 & --                         & 16.06$\,\pm\,$0.47  &   --               & 15.96$_{-0.68}^{+0.71}$$_{-0.3}^{+0.5}$  \\\hline
 7.0 -- 9.0 &  7.90 & 16.7$\,\pm\,2.5\,\pm\,$0.4 &  --                 &   --               & 16.03$_{-0.67}^{+0.69}$$_{-0.3}^{+0.5}$  \\
 8.0 -- 9.0 &  8.48 & --                         & 16.73$\,\pm\,$0.60  &   --               & 16.12$_{-0.65}^{+0.68}$$_{-0.3}^{+0.5}$  \\
 8.0 -- 10.0&  8.87 & --                         &  --                 & 16.27$\,\pm\,$0.37 & 16.17$_{-0.65}^{+0.67}$$_{-0.3}^{+0.5}$  \\
 9.0 -- 10.0&  9.48 & --                         & 18.53$\,\pm\,$0.55  &   --               & 16.25$_{-0.64}^{+0.65}$$_{-0.3}^{+0.5}$  \\
10.0 -- 11.0& 10.48 & --                         & 18.66$\,\pm\,$0.76  &   --               & 16.36$_{-0.62}^{+0.63}$$_{-0.3}^{+0.5}$  \\
10.0 -- 12.0& 10.90 & --                         &  --                 & 16.76$\,\pm\,$0.47 & 16.4 $_{-0.61}^{+0.63}$$_{-0.3}^{+0.5}$  \\\hline
11.0 -- 12.0& 11.49 & --                         & 19.16$\,\pm\,$0.78  &   --               & 16.45$_{-0.61}^{+0.62}$$_{-0.3}^{+0.5}$  \\
12.0 -- 13.5& 12.71 & --                         & 17.5$\,\pm\,$1.1    &   --               & 16.55$_{-0.59}^{+0.60}$$_{-0.3}^{+0.5}$  \\
12.0 -- 14.0& 12.91 & --                         &  --                 & 15.95$\,\pm\,$0.65 & 16.57$_{-0.59}^{+0.60}$$_{-0.3}^{+0.5}$  \\
13.5 -- 15.0& 14.22 & --                         & 19.8$\,\pm\,$1.2    &   --               & 16.65$_{-0.58}^{+0.58}$$_{-0.3}^{+0.5}$  \\
14.0 -- 17.0& 15.31 & --                         &  --                 & 17.83$\,\pm\,$0.67 & 16.72$_{-0.57}^{+0.57}$$_{-0.3}^{+0.5}$  \\
15.0 -- 17.0& 15.95 & --                         & 20.8$\,\pm\,$1.2    &   --               & 16.75$_{-0.56}^{+0.57}$$_{-0.3}^{+0.5}$  \\\hline
17.0 -- 20.0& 18.40 & --                         & 22.0$\,\pm\,$1.3    & 17.06$\,\pm\,$0.97 & 16.86$_{-0.55}^{+0.55}$$_{-0.3}^{+0.5}$  \\
20.0 -- 25.0& 22.28 & --                         & 24.5$\,\pm\,$1.8    & 17.46$\,\pm\,$1.10 & 16.99$_{-0.53}^{+0.53}$$_{-0.3}^{+0.5}$  \\
25.0 -- 30.0& 27.31 & --                         & 18.1$_{-4.0}^{+3.3}$& 16.98$\,\pm\,$1.94 & 17.11$_{-0.51}^{+0.51}$$_{-0.3}^{+0.5}$  \\
30.0 -- 40.0& 34.36 & --                         & 28.5$_{-4.5}^{+3.9}$& 17.95$\,\pm\,$2.26 & 17.23$_{-0.49}^{+0.49}$$_{-0.3}^{+0.5}$  \\
     --     & 112   & --                         &                     & 20.16$\,\pm\,$1.65 & 17.63$_{-0.43}^{+0.42}$$_{-0.3}^{+0.5}$  \\
\end{tabular}
\end{ruledtabular}
\end{center}
\end{table*}
%%%%%%%%%%%%%%%%%%%%%%%%%%%%%%%%%%%%%%%%%%%%%%%%%%%%%%%%%%%%%%%%%%%%%%%

As one can see from the right panel of Fig.\ \ref{fig:data-ellipses},
the situation changes considerably when the high-$Q^2$ tail of the
BaBar data \cite{BaBar09} is included in the data analysis.
Using the same designations as in the left panel, we show the analogous
unified error ellipses, observing that now the ellipsoid has no
intersection with the $(a_2,a_4)$ plane.
Moreover, the composed error ellipse resulting from the 2D analysis
(solid blue line) moves out of the region of the negative values of
$a_4$ and inside its positive domain.
This is accompanied by the significantly worse value
$\chi^2_\text{ndf}\approx 2$,
in sharp contrast to the value $\chi^2_\text{ndf}\approx 0.5$
we found for the data set in the CLEO-regime.
On the other hand, the unified $1\sigma$ error ellipse of the 3D
projections on the $(a_2,a_4)$ plane (larger dashed red ellipse),
keeps its position unchanged and still harbors a big portion of the
area for the $a_2$, $a_4$ values, compatible with the results from
NLC SRs (shaded rectangle in green color).

Comparison with the lattice findings shows that the 3D error ellipse
lies almost entirely within the boundaries from \cite{Lat06}
(dashed vertical lines)
but intersects with the interval determined in \cite{Lat10}
(dashed-dotted vertical lines)
only for the larger values of $a_2$.
The analogous ellipse of the 2D analysis only poorly complies with
this small $a_2$ window of \cite{Lat10}, whereas it partly overlaps
with the low end of the $a_2$ range determined in \cite{Lat06}.
Thus, the mutual agreement between the 2D and the 3D analysis,
observed in the left panel, now deteriorates.
Moreover, it becomes obvious from this figure that Model III from
\cite{ABOP10} has a projection on the $(a_2,a_4)$ plane that lies
outside of all considered $1\sigma$ error ellipses of the data.
Note incidentally that selecting for the Borel parameter
the value $M^2=1.5$~GeV$^2$, as used in \cite{ABOP10},
we find for their Model III an agreement with
the BaBar data on the level of $\chi^2_\text{ndf}\gtrsim 1.5$.

\begin{figure}[b!]\vspace*{-3mm}
\centerline{\includegraphics[width=0.48\textwidth]{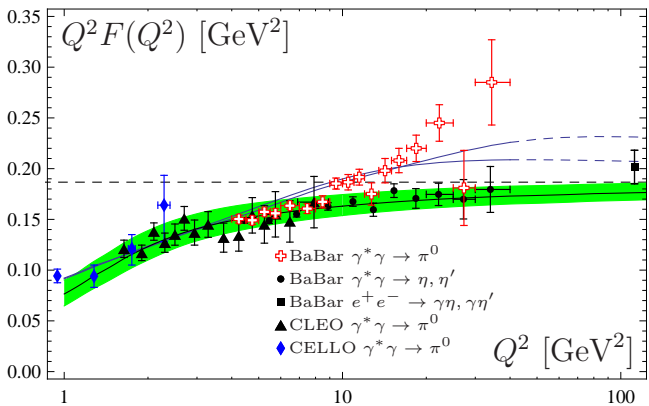}}
\vspace*{-3mm}
\caption{(color online). Predictions for the scaled form factor
$Q^2F^{\gamma^*\gamma\pi}(Q^2)$ using a logarithmic scale for $Q^2$.
The predictions using the BMS ``bunch'' of $\pi$ DAs are displayed
in the form of a shaded (green) strip,
with the solid line denoting the BMS model \protect\cite{BMS01}.
The width of the strip contains various theoretical uncertainties
explained in the text.
The two single solid (blue) lines reproduce the predictions
of Agaev et al.\ \protect\cite{ABOP10} with their models I and III,
whereas the prediction using their model II---not shown---would be
in-between.
[For the sake of comparison, we extended these predictions
beyond 40~GeV$^2$ in the form of dashed lines up to the remote point
of 112~GeV$^2$].
The experimental data are taken from various experiments
\protect\cite{CELLO91,CLEO98,BaBar09,BaBar11-BMS,Aub2006}.
\label{fig:FF-predictions}}
\end{figure}

This discussion becomes more substantiated by inspecting
Fig.\ \ref{fig:FF-predictions} which compares theoretical predictions
with various experimental data with appropriate designations
provided in the figure.
Our calculations are displayed in the form of a shaded strip
(green color) with a width reflecting various theoretical
uncertainties.
These are (i) uncertainties owing to the spread of the ``bunch'' of
the pion DAs derived in \cite{BMS01},
(ii) those originating from the variation
of the twist-four parameter $\delta^2$,
and
(iii) those related to the sum of the NNLO$_{\beta_0}$ term
and the twist-six contribution \cite{ABOP10}.
Because these two latter contributions are comparable in size, they
almost mutually cancel.
For a more detailed understanding, we have compiled the calculated
form-factor values for the BMS DA at each measured point in
Table \ref{tab:ff-values-table-1} (last column),
including also the aforementioned theoretical
errors.
Note that the first error originates from items
(i) and (ii), whereas the second one stems from item (iii).

%%%%%%%%%%%%%%%%%%%%%%%%%%%%%%%%%%%%%%%%%%%%%%%%%%%%%%%%%%%%%%%%%%%%%%% FIGURE 4
%%%%%%%%%%%%%%%%%%%%%%%%%%%%%%%%%%%%%%%%%%%%%%%%%%%%%%%%%%%%%%%%%%%%%%%

As one clearly sees from Fig.\ \ref{fig:FF-predictions}, these
predictions (shaded strip in green color), obtained with LCSRs within
the standard collinear factorization scheme of QCD, cannot reproduce
the significant rise of the high-$Q^2$ tail of the
$\gamma^*\gamma\to \pi^0$ BaBar data \cite{BaBar09}.
But, surprisingly, they are in very good agreement with the BaBar data
for the two-photon $\eta$ and $\eta^\prime$ decays \cite{BaBar11-BMS}
using the description of the $\eta-\eta^{\prime}$ mixing in the quark
flavor basis \cite{FKS98}.
These data are also shown in the fifth column of
Table \ref{tab:ff-values-table-1}.
Under the assumption that the state
$|n\rangle = (|\bar{u}u\rangle + |\bar{d}d\rangle)/\sqrt{2}$
has a DA similar to that of the pion, one can link the
$\gamma^*\gamma\to |n\rangle$ transition form factor,
multiplied by $3/5$, to the form factor $\gamma^*\gamma\to\pi^0$,
where the prefactor arises from the quark charges.
Strictly speaking, one should compare the state $\pi^0$ not
with the physical particles $\eta$ and $\eta^\prime$ but with the
other neutral states $\eta_1$ and $\eta_8$ of the pseudoscalar
meson sector.
The observed discrepancy of the two BaBar data sets indicates that
the structure of $\eta_8$ is very different from that of $\pi^0$,
which looks implausible in view of the fact that both these states
belong to the octet (a detailed discussion of this issue is given in
\cite{Kro10sud} and \cite{BCT11}).
Whether this is a true dynamical effect or accidental cannot be
answered rigorously at present.
One realizes from this figure that two BaBar data points for the
$\gamma^*\gamma\to\pi^0$ measurement \cite{BaBar09}
at $Q^2=12.7$~GeV$^2$ and $27.3$~GeV$^2$
also lie inside the shaded strip, being perfectly in line with the data
for the $|n\rangle$ form factor.
Moreover, our predictions (shaded strip) approach, together with these
data, the asymptotic QCD limit $\sqrt{2}f_\pi$ from below.
In marked contrast, the predictions derived by Agaev et al.
\cite{ABOP10}---single solid lines in this figure---fail to comply with
these data and exceed considerably the asymptotic QCD limit.
Unfortunately, they even fail to provide agreement with the
$\gamma^*\gamma\to \pi^0$ BaBar data as well.
Indeed, as one sees from Fig.\ \ref{fig:FF-predictions},
the characteristic rise of these data with $Q^2$ beyond 10~GeV$^2$
cannot be reproduced, mainly because the enhancement provided
by a large and positive coefficient $a_4$ is not really
sufficient---even augmenting it with the inclusion
of some more higher coefficients ${a_6, \ldots}$---so that the form
factor gradually flattens out instead of increasing with $Q^2$.
However, in view of the fluctuations of the BaBar data
at higher $Q^2$ values, one cannot rule out a flatter
behavior than the fit in Eq.\ (\ref{eq:power}) proposed by BaBar.
To quantify the above statements and facilitate comparisons with
other approaches, we collect in Table \ref{tab:chi.Model}
the statistical properties pertaining to the data fits,
using various sets of data and selected pion DA models.
Let us also mention that the variation of the lower
bound of $Q^2$ in the statistical analysis, e.g., the exclusion of the
lowest six experimental points to prevent a strong influence
of twist-4 uncertainty, does not affect the pattern of
the $\chi_\text{ndf}^2$ values of the BMS DA, Asy DA and the CZ DA,
shown in Table Table \ref{tab:chi.Model}.
A detailed analysis of the variation of the lower bound in $Q^2$ has
been given in \cite{BMS03}.

Be that as it may, the most immediate conclusion is that the BaBar
data for the pion and the $|n\rangle$ state seem antithetical and
cannot be equally reproduced within the standard QCD scheme.

%%%%%%%%%%%%%%%%%%%%%%%%%%%%%%%%%%%%%%%%%%%%%%%%%%%%%%%%%%%%%%%%%%%%%%%
%%%%%%%%%%%%%%%%%%%%%%%%%%%%%%%%%%%%%%%%%%%%%%%%%%%%%%%%%%%% TABLE II
%%%%%%%%%%%%%%%%%%%%%%%%%%%%%%%%%%%%%%%%%%%%%%%%%%%%%%%%%%%%%%%%%%%%%%%
\begin{table*}[t!]
\begin{center}
\begin{ruledtabular}
\caption{Statistical properties of selected pion DA models (first column)
with corresponding coefficients $a_n$ (second column)
used in the calculation of the pion-photon transition form factor by means
of LCSRs.
The last two columns show the values of
$\chi_\text{ndf}^2\equiv\chi^2/{\rm ndf}$
(with ndf~$=$~number of degrees of freedom)
for the data in the CLEO regime and for the whole set
of the data, respectively.
\label{tab:chi.Model}}
\smallskip
\begin{tabular}{lccc}
 $\pi$ DA model                     & $(a_2, a_4, \ldots )_{\mu^2=\mu_\text{SY}^2}$         & $\chi_\text{ndf}^2$ (1--9~GeV$^2$)
                                                                                            & $\chi_\text{ndf}^2$ (1--40~GeV$^2$)
                                                                                                            \\\hline
 BMS DA ({\footnotesize\ding{54}})  & $(0.141, -0.089)$                                     & $0.5$ & $3.1$ \\
 Asy DA ({\footnotesize\ding{117}}) & $(0, 0)$                                              & $4.7$ & $7.9$ \\
 CZ DA ({\footnotesize\ding{110}})  & $(0.394, 0)$                                          & $32.3$& $25.5$\\
 Model I \protect\cite{ABOP10}      & $(0.084, 0.137, 0.088, 0.063, 0.048, 0.039)$ & $2.8$ & $2.4$\\
 Model III \protect\cite{ABOP10} ({\footnotesize\ding{115}})
                                    & $(0.104, 0.123, 0.039)$                      & $3.2$ & $2.8$
\end{tabular}
\end{ruledtabular}
\end{center}\vspace*{-3mm}
\end{table*}
%%%%%%%%%%%%%%%%%%%%%%%%%%%%%%%%%%%%%%%%%%%%%%%%%%%%%%%%%%%%%%%%%%%%%%%

%%%%%%%%%%%%%%%%%%%%%%%%%%%%%%%%%%%%%%%%%%%%%%%%%%%%%%%%%%%%%%%%%%%%%%% FIGURE 5
\begin{figure}[t!]\vspace*{-3mm}
\centerline{\includegraphics[width=0.48\textwidth]{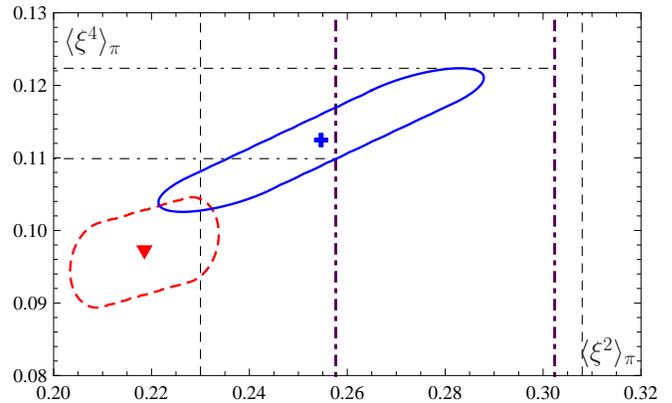}}
 \vspace*{-3mm}
 \caption{(color online). Predictions for the moments
  $\langle \xi^2\rangle_\pi$ and
  $\langle \xi^4\rangle_\pi$
  at the lattice scale $\mu^2_\text{Lat}=4$~GeV$^2$.
  The solid (blue) ellipse corresponds to our choice of $M^2$,
  whereas the dashed (red) one results when using $M^2=1.5$~GeV$^2$.
  The vertical lines show the range of values
  computed on the lattice:
  dashed line---\protect\cite{Lat06};
  dashed-dotted (violet) line---\protect\cite{Lat10}.
 \protect\label{fig:Moments}}
\end{figure}
%%%%%%%%%%%%%%%%%%%%%%%%%%%%%%%%%%%%%%%%%%%%%%%%%%%%%%%%%%%%%%%%%%%%%%%

The accuracy of the experimental data on $\gamma^*\gamma\to \pi^0$
and the precision of current lattice simulations in extracting
constraints on the second moment
$\langle \xi^2 \rangle_\pi$
of the pion DA have both reached a level that can be used to determine
the range of values of the next moment
$\langle \xi^4 \rangle_\pi$,
where both experimental data and lattice computations overlap.
This possibility was first pointed out in \cite{Ste08} and the
following range of values was extracted from the $1\sigma$
error ellipse of the CLEO data \cite{CLEO98} in conjunction
with the lattice constraints for
$\langle \xi^2 \rangle_\pi$ from \cite{Lat07}:
$\langle \xi^4 \rangle_\pi \in [0.095 \div 0.134]$ at
$\mu^2_\text{Lat}=4$~GeV$^2$ and for $M^2=0.7$~GeV$^2$.
Here we refine this procedure in the following way: first,
we map the $(a_2,a_4)$ plane
onto the
$(\langle \xi^2 \rangle_\pi,\langle \xi^4 \rangle_\pi)$ plane.
Then, we combine the mapped $1\sigma$ error ellipse
(area enclosed by a solid blue line)
in Fig.\ \ref{fig:Moments} for the data sets in the CLEO regime
(cf.\ Fig.\ \ref{fig:data-ellipses})
with the constraints from \cite{Lat06} and \cite{Lat10}
(Table XII there),
the aim being to extract that range of values of
$\langle \xi^4 \rangle_\pi$
where they overlap.
The corresponding results at the typical lattice scale
$\mu^2_\text{Lat}=4$~GeV$^2$
are, respectively,
(i) $\langle \xi^2 \rangle_\pi \in [0.23 \div 0.29]$
and
$\langle \xi^4 \rangle_\pi \in [0.102 \div 0.122]$,
(ii)
$\langle \xi^2 \rangle_\pi \in [0.26 \div 0.29]$
and
$\langle \xi^4 \rangle_\pi \in [0.11 \div 0.122]$.
These windows were extracted by employing a $Q^2$-dependent
Borel parameter, as before in this analysis.
Using instead $M^2=1.5$~GeV$^2$, as in \cite{ABOP10}, one would obtain
only a small intersection of the corresponding region,
shown in Fig.\ \ref{fig:Moments} by the dashed (red) line,
with the lattice constraints of~\cite{Lat06}, amounting to the value
$\langle \xi^4 \rangle_\pi \simeq0.1$,
while there would be no intersection at all with the lattice estimates
of \cite{Lat10}.
The sensitivity of $\langle \xi^2 \rangle_\pi$ on the choice of the
Borel parameter $M^2$ provides a handle to consider the existing
lattice computations \cite{Lat06,Lat10} as providing independent
evidence in support of the original prescription for the Borel
parameter in the LCSRs for the pion-photon transition
form factor \cite{Kho99,SY99}.

\section{Summary and conclusions}
\label{sec:concl}

In summary, three different sets of experimental data on the
pion-photon transition form factor have been analyzed with the help of
QCD LCSRs along the lines of our earlier works in \cite{BMS02,BMS03}
and adopting the reasoning expressed in \cite{Kho99,SY99}.
The following key ingredients have been taken into account:
(i) the NLO radiative correction,
(ii) the twist-four contribution,
and
(iii) the NNLO$_{\beta_0}$ QCD correction \cite{MMP02,MS09} together
with the recently computed \cite{ABOP10} twist-six contribution,
these two by means of theoretical uncertainties.
This is possible in our approach because in the range of the
Borel parameter adopted,
$M^2<1.0$~GeV$^2$,
these two corrections, though with opposite signs, have almost the
same absolute (not large) magnitude.
Carrying out the analysis, evidence has been obtained that the data
from CELLO \cite{CELLO91}, CLEO \cite{CLEO98}, and BaBar \cite{BaBar09}
between 1 and 9~GeV$^2$ favor a pion distribution amplitude
with endpoint suppression, like the BMS model \cite{BMS01}
(see the left panels of Fig.\ \ref{fig:3D-graphs}
 and Fig.\ \ref{fig:data-ellipses}).
The key for the overlap between the error ellipses of the data
in the $(a_2,a_4)$ plane with the allowed region from NLC SRs
is a \emph{negative} value of the Gegenbauer coefficient
$|a_4|\lesssim a_2$.
Beyond 10~GeV$^2$, the best fit to the BaBar data on
$F^{\gamma^*\gamma\to\pi^0}(Q^2)$
requires a sizeable coefficient $a_6$ comparable with the lower
coefficients (cf.\ right panel of Fig.\ \ref{fig:3D-graphs}).

The comparison of our theoretical predictions for the scaled
form factor
$Q^2F^{\gamma^*\gamma\pi}(Q^2)$
in Fig.\ \ref{fig:FF-predictions}
makes it apparent that there is an antithetic trend between the
BaBar data on the pion-gamma transition form factor \cite{BaBar09}
and those extracted from the the
$\gamma^*\gamma\to\eta(\eta^\prime)$
transition form factors \cite{BaBar11-BMS}.
While the latter agree very well with the BMS predictions
(solid line in Fig.\ \ref{fig:BaBar-eta-fig18} and
shaded strip in Fig.\ \ref{fig:FF-predictions}),
the high-$Q^2$ tail of the $\pi^0$ BaBar data requires a pion DA with
sizeable (or even growing with $n$) higher Gegenbauer coefficients
$a_n$, or alternative theoretical schemes outside the standard QCD
factorization approach,
see, e.g., \cite{Rad09,Pol09,Dor09,KOT10plb,SZ11}.
Similar conclusions were also drawn in \cite{RRBGGT10}
using Dyson--Schwinger equations.

This intriguing behavior of the BaBar data cannot be reconciled
with QCD factorization in combination with LCSRs,
despite the opposite claims expressed in \cite{ABOP10}.
As one sees from Fig.\ \ref{fig:FF-predictions}, this is not possible
for our choice of the Borel window $M^2\approx0.7-0.9$~GeV$^2$,
giving rise to the shaded strip, but also not for the higher values
$M^2=1.5\pm 0.5$~GeV$^2$ employed in \cite{ABOP10} (single solid lines
in the same figure).
Even a large positive coefficient $a_4$ and a sizeable value of $a_6$
cannot provide sufficient reinforcement of the
$\gamma^*\gamma\to\pi^0$ form factor to bridge the gap to the
high-$Q^2$ BaBar data and reproduce their increase with $Q^2$
{without the loss of the statistical accuracy.
Quantitatively, this means that involving in the 3D data analysis
the high-$Q^2$ tail of the BaBar data, makes the description inevitably
worse ($\chi^2_\text{ndf}\geq 1$) in comparison with the analysis
applicable to the CLEO regime ($\chi^2_\text{ndf} \approx 0.4$).
Remarkably, also the recent light-front holographic analysis in
Ref.\ \cite{BCT11} yields predictions which are incompatible with
the strong rise of the BaBar data at high $Q^2$, while being in
agreement with those extracted from the
$\gamma^*\gamma\to\eta(\eta^\prime)$ transition form factors and
rather close to our results in this region.

Bottom line:
New measurements, e.g., by the Belle Collaboration, may help resolving
the controversies around the BaBar data and understanding
the dichotomy between the pion and the state $|n\rangle$
which indicates a strong violation of flavor symmetry.
In addition, a precise lattice estimate of $\langle \xi^4 \rangle_\pi$
could be instrumental in fixing the sign of the Gegenbauer coefficient
$a_4$ of the pion DA.
Moreover, our detailed predictions in Table
\ref{tab:ff-values-table-1} in conjunction with Fig.\
\ref{fig:FF-predictions} may serve as a sort of `reference model' for
various purposes: theoretical and experimental.

%%%%%%%%%%%%%%%%%%%%%%%%%%%%%%%%%%%%%%%%%%%%%%%%%%%%%%%%%%%%%%%%%%%%%%%
\acknowledgments
%%%%%%%%%%%%%%%%%%%%%%%%%%%%%%%%%%%%%%%%%%%%%%%%%%%%%%%%%%%%%%%%%%%%%%%
We would like to thank Anatoly Efremov, Simon Eidelman, Andrei Kataev,
and Dmitri Naumov for stimulating discussions and useful remarks.
A.B., S.M., and A.P. are thankful to Prof. Evgeny Epelbaum and
Prof. Maxim Polyakov for the warm hospitality at Bochum University,
where the major part of this investigation was carried out.
A.B., S.M., and A.P. acknowledge financial support from Nikolay Rybakov.
A.P. also wishes
to thank the Ministry of Education and Science of the Russian Federation
(``Development of Scientific Potential in Higher Schools''
 projects No.\ 2.2.1.1/12360 and No.\ 2.1.1/10683).
This work was supported in part by the Heisenberg--Landau Program under
Grant 2011, the DAAD (A.P.), the Russian Foundation for Fundamental
Research (Grant No.\ 09-02-01149), and the BRFBR--JINR Cooperation
Program under contract No.\ F10D-002.

%%%%%%%%%%%%%%%%%%%%%%%%%%%%%%%%%%%%%%%%%%%%%%%%%%%%%%%%%%%%%%%%%%%%%%%%%
%% BibTeX Commands %%%%%%%%%%%%%%%%%%%%%%%%%%%%%%%%%%%%%%%%%%%%%%%%%%%%%%
%%%%%%%%%%%%%%%%%%%%%%%%%%%%%%%%%%%%%%%%%%%%%%%%%%%%%%%%%%%%%%%%%%%%%%%%%
%%\bibliographystyle{apsrev}
%%\bibliographystyle{apsrev-ab}
%\bibliographystyle{prsty-ab}
%\bibliography{pion,lambda,nonloc}
%\end{document}
%%%%%%%%%%%%%%%%%%%%%%%%%%%%%%%%%%%%%%%%%%%%%%%%%%%%%%%%%%%%%%%%%%%%%%%%%

\newcommand{\noopsort}[1]{} \newcommand{\printfirst}[2]{#1}
  \newcommand{\singleletter}[1]{#1} \newcommand{\switchargs}[2]{#2#1}

\end{document}